\documentclass[conference]{IEEEtran}
\IEEEoverridecommandlockouts
\usepackage{cite}
\usepackage{amsmath,amssymb,amsfonts}
\usepackage{graphicx}
\usepackage{textcomp}
\usepackage{xcolor}
\def\BibTeX{{\rm B\kern-.05em{\sc i\kern-.025em b}\kern-.08em
    T\kern-.1667em\lower.7ex\hbox{E}\kern-.125emX}}

\makeatletter
\newcommand{\linebreakand}{%
  \end{@IEEEauthorhalign}
  \hfill\mbox{}\par
  \mbox{}\hfill\begin{@IEEEauthorhalign}
}
\makeatother
\usepackage{algorithm,algpseudocode} 
\begin{document}

\title{PRIVACY AWARE MEMORY FORENSICS}
% \author{
%   \IEEEauthorblockN{1\textsuperscript{st} Janardhan Kalikiri}
%   \IEEEauthorblockA{\textit{Indian Institute of Technology, Jammu}\\
%   jana1tech@gmail.com}
%   \and
%   \IEEEauthorblockN{2\textsuperscript{nd} Gaurav Varshney}
%   \IEEEauthorblockA{\textit{Indian Institute of Technology, Jammu}\\
%   gaurav.varshney@iitjammu.ac.in}
%   \and
%   \and
%   \IEEEauthorblockN{3\textsuperscript{rd} Jaswinder Kour}
%   \IEEEauthorblockA{\textit{Indian Institute of Technology, Jammu}\\
%   2021rcs2010@iitjammu.ac.in}
%   \and
%   \IEEEauthorblockN{4\textsuperscript{th} Tarandeep Singh}
%   \IEEEauthorblockA{\textit{Indian Institute of Technology, Jammu}\\
%   tarandeep42@gmail.com}
% }

\author{
  \IEEEauthorblockN{1\textsuperscript{st} Janardhan Kalikiri}
  \IEEEauthorblockA{\textit{Indian Institute of Technology, Jammu}\\
  jana1tech@gmail.com}
  \and
  \IEEEauthorblockN{2\textsuperscript{nd} Gaurav Varshney}
  \IEEEauthorblockA{\textit{Indian Institute of Technology, Jammu}\\
  gaurav.varshney@iitjammu.ac.in}
  \linebreakand
  \IEEEauthorblockN{3\textsuperscript{rd} Jaswinder Kour}
  \IEEEauthorblockA{\textit{Indian Institute of Technology, Jammu}\\
  2021rcs2010@iitjammu.ac.in}
  \and
  \IEEEauthorblockN{4\textsuperscript{th} Tarandeep Singh}
  \IEEEauthorblockA{\textit{Indian Institute of Technology, Jammu}\\
  tarandeep42@gmail.com}
}

\maketitle
\begin{abstract} In recent years, insider threats and attacks have been increasing in terms of frequency and cost to the corporate business. The utilization of end-to-end encrypted instant messaging applications (WhatsApp, Telegram, VPN) by malicious insiders raised data breach incidents exponentially. The Securities and Exchange Board of India (SEBI) investigated reports on such data leak incidents and reported about twelve companies where earnings data and financial information were leaked using WhatsApp messages. Recent surveys indicate that 60\% of data breaches are primarily caused by malicious insider threats. Especially, in the case of the defense environment, information leaks by insiders will jeopardize the country’s national security. Sniffing of network and host-based activities will not work in an insider threat detection environment due to end-to-end encryption. Memory forensics allows access to the messages sent or received over an end-to-end encrypted environment but with a total compromise of the user's privacy. In this research, we present a novel solution to detect data leakages by insiders in an organization. Our approach captures the RAM of the insider’s device and analyses it for sensitive information leaks from a host system while maintaining the user's privacy. Sensitive data leaks are identified with context using a deep learning model. The feasibility and effectiveness of the proposed idea have been demonstrated with the help of a military use case. The proposed architecture can however be used across various use cases with minor modifications. 
\end{abstract}


\section{Introduction}

Insider threats come from users with legitimate authorized access to the organization’s assets and who exploit them intentionally or accidentally. The 2021 insider threat report by Cybersecurity Insider states that 98\% of organizations feel vulnerable to insider threats. Most organizations (85\%) consider unified visibility and access control across all apps, devices, web destinations, on-premises resources and infrastructure as significant to moderately important to prevent insider threats cite{survey}.In many insider trading incidents, the violator posted information about the financial performance indicators of companies before their official disclosure through WhatsApp Messenger. This significantly impacted the stock price
and gave the dishonest brokers a competitive advantage. The regulatory authorities could not detect these leaks due to the lack of monitoring technologies for these IM applications, which have become a channel for sensitive data leakages.

\par   SolarWinds Data Loss Prevention with ARM, Trustifi Outbound Shield, Manage Engine Endpoint DLP Plus are commonly used data loss prevention tools in companies but
none of them deals with the prevention of data leaks through IM apps\cite{comparitech}.  In this paper, we for the very first time address this issue and propose an effective methodology to prevent insider attacks using memory forensics while ensuring user privacy. We are trying to demonstrate a way through which live memory forensics-based insider threat detection solutions can run on end systems with an assurance that user's privacy won't be hampered.

\par 
Memory is the workplace of the processor. Therefore, Memory Forensics is one of the most effective digital forensic disciplines that aims to extract digital evidence from volatile data existing in the RAM\cite{generic}.  However one of the challenges in this process is that the RAM also contains private/personal data of the user/employee such as chat with family or other information related to his/her access to social media accounts etc. A full analysis of RAM by the employer will compromise the privacy of the user. In this research work, we focus on how an employer can have access to meaningful traces and indicators of data leakage from the end systems yet he can assure a benign user that his personal data privacy stakes are not affected. We propose a novel idea of a privacy-aware memory forensics framework that solves this problem. This paper is organized as follows: Section II discusses related work in the field of memory forensics and user privacy preservation and identified research gaps. Section III provides a detailed explanation
of our proposed work. In section IV, the implementation of our approach with experiments, results, and their analysis is elucidated. Finally, in section V, we conclude and throw light on future work in this direction.

\section{Related Work}

 Memory forensics can be defined as the forensics field of computer science, which is growing very fast in assisting the investigator in investigating malicious activities [11]. Memory forensics evolved in 2004 and was introduced by Michael Ford. Many tools were later developed in this area to help forensics analysis of memory including Responder PRO, Memoryze, MoonSols Winodws Memory Toolkitwinen, Belkasoft Live RAM Capturer, etc.; Memory forensic is now being used actively in forensics evidence collection and in real-time
incident response.
 \par  S. Srinivasan \cite{policy} proposed the
 privacy preservation methodology in a digital investigation by implementing policies while handling user data but did not implement the proposed design. Frank Y.W. Law et al.\cite{Enc} proposed a searchable encryption model to provide privacy in the digital investigation model where the disk image is analyzed. It didn't include the contextual aspect of privacy preservation. M. Burmester et al.\cite{choose} proposed policy-based privacy preservation in disk forensics. The search performed was keyword-based based thus could include the user's private data. Waleed Halboob et al. \cite{quarter} proposed a concept of quaternary privacy levels in computer forensics using the investigators' user data access control rights. 
\par Majority of existing published digital forensics investigation models or procedures have not incorporated the strategy for supporting data privacy protection \cite{email}, \cite{chat}, \cite{network}, \cite{doh} and \cite{choose}. 
Also the previous works on applications of memory forensics to solve various problems\cite{email}, \cite{chat} and \cite{network} are computationally expensive due to the analysis of the entire RAM for their solution. The comparison of previous research work is tabulated in Table I. The following research gaps are identified:
\begin{itemize}
    \item Most of the schemes have focused on efficient memory forensics practices that require less computation.

    \item None of the solutions are focusing on live memory forensics over instant messaging applications.
    \item It was identified that none of the schemes focus on the user privacy aspects of memory forensics.
    \item Not many schemes have focused on context-based searches for live memory forensics.
\end{itemize}
These gaps have motivated our research on privacy-aware memory forensics. We propose a novel malicious insider detection scheme. We have taken WhatsApp as the IM application for our research and development which uses a BERT pre-trained model over sensitive context for the detection of sensitive messages from the WhatsApp memory dump acquired live from the RAM of a desktop.
The proposed method performs a context-based sensitive WhatsApp messages detection on the Windows 10 operating system.
   
The major contributions of this paper are:
\begin{itemize}
    \item Detailed study of Memory Forensics and issues concerning user privacy. 
    \item Designing an algorithm that can capture per-process live memory and individual chats of an instant messaging application (WhatsApp).
    \item Proof of concept sensitive message detection tool over Windows.
    \item Military sensitive training data set for BERT model for detecting malicious insiders attack in the military use case.   
\end{itemize}
\begin{table*}[h!]
\caption{Comparison of previous work }\label{tab1}
\large
\begin{center}
\resizebox{\textwidth}{!}{\begin{tabular}{|l|l|l|l|l|l|l|}
\hline
\textbf{Author} & \textbf{Work} & \textbf{Memory } & \textbf{Privacy Preservation }  & \textbf{Lightweight } & \textbf{Keyword} & \textbf{Implemented}\\
 & \textbf{Performed} & \textbf{RAM/Disk} & \textbf{in Mem. Forensics} & \textbf{or Not} & \textbf{/Context} & / \textbf{Proposed} \\
\hline
Srinivasan \cite{policy} & Policies discussed  & Disk & No & No & NA & Proposed \\
\hline
Waleed Halboob et al.\cite{quarter} & Quaternary   & Disk & No & No & NA & Proposed \\
&  Privacy level &  & & & &\\
\hline
M Burmester et al.\cite{choose} & Cryptographic  & Disk & No & No & Keyword & Implemented \\
In this paper, we propose&  Techniques &  & & & &\\
\hline
A Nieto, R Rios\cite{Psurvey} & Privacy-aware  & Disk & No & No & NA & survey \\
&  digital forensics &  & & & &\\
\hline
Ali Dehghantanha et al. \cite{privacy}& Privacy issues   & Disk & No & No & NA & survey \\
&  in investigation &  & & & &\\
\hline
Padmavathi Iyer et al. \cite{email}& Detection of   & Live & No & No & Keyword & Implemented \\
 & email spoofing &  & & & &\\
\hline
Abdullah Kazim et al. \cite{chat}& Recovery of chat   & Live & No & No & Keyword & Implemented \\
 & messages \&  &  & & & &\\
 & encryption key &  & & & &\\
\hline
Al-Saleh et al. \cite{network} & Network   & Live & No & No & NA & Implemented \\
& Reconnaissance &  & & & &\\
\hline

Gaurav et al. \cite{doh} & Detection of    & Live & No & No & Keyword & Implemented \\
 &  URLs visited Via DoH &  & & & &\\
\hline

\end{tabular}}
\end{center}
\end{table*}

\section{Proposed Work}
This paper proposes an NLP-based architecture for Insider's attack to ensure user privacy in memory forensics. We demonstrate a novel approach for detecting sensitive data leaks by an insider (intentional or accidental) using the Windows desktop instant messaging WhatsApp while preserving user privacy in a defense use case. The goal is achieved in two significant steps: firstly the WhatsApp chats matching the desired context are retrieved to identify sensitive messages, and in the second step an alert message is generated and sent to the security center if any sensitive data leak is identified, appending the login user’s mobile number for further action.
The proposed architecture is illustrated in Figure 1.

\begin{figure}
\centering
    \includegraphics[ width=.35\textwidth]{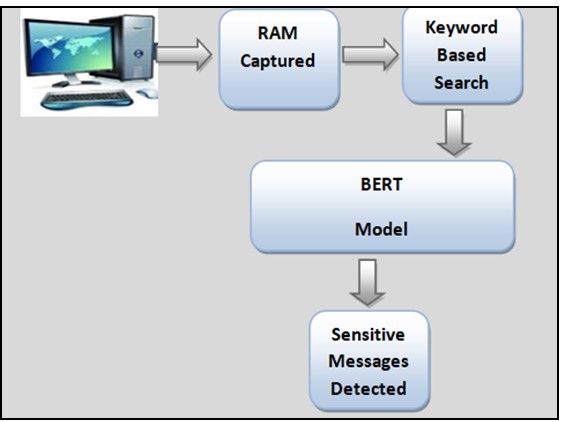}
    \caption{Sensitive Data Detection Model}
    \label{Fig:1}
\end{figure}
\subsection{Data Set Generation}
Due to the non-availability of the military data set consisting of sensitive data for fine-tuning the BERT module, we generated our training data set manually. For pre-training of the BERT model, we can train the model with a manually generated small data set for efficient performance. Post-analysis of recent defense data leakage cases and the following topics are considered while developing the data set. 

\begin{itemize}
    \item Modernisation plans of defense equipment  
    \item Combat capabilities of the armed forces
    \item Serviceability of weapon and mission-critical equipment
    \item Movement of the fleet, battalion, military forces, and VIPs
    \item Information about special operations
\end{itemize}   

\subsection{Live Memory Capturing: Using ProcDump}
In our model, a Microsoft command-line utility called \textbf{ProcDump} \cite{proc} captures a particular process (WhatsApp) instead of the entire RAM (in GB). The proposed model hence is lightweight and utilizes significantly low computation and storage. ProcDump required administrative-level rights to dump a process memory. In most of the previous work on memory forensics, third-party tools are used to capture volatile data. The entire RAM was inspected for conducting the experiment. In our case, we only dump the WhatsApp process memory thereby reducing the dump storage requirement and improving the privacy from the first step of our analysis.   
With the specifications given in Table II, we implemented our privacy-aware memory forensics architecture on a user machine to detect sensitive data leaks over WhatsApp Desktop app installed on a Windows Machine. Our implementation over WhatsApp is a proof of our concept that privacy-aware memory forensics can be done over IM applications. To implement this our first step was to capture the memory of the WhatsApp desktop application in real-time. We send sample sensitive data on individual and group chats along with personal chat messages from systems/mobiles. WhatsApp desktop application launches multiple (generally 6 to 7) background processes while running on the user machine. All processes do not contain chat messages, thus capturing correct process memory with chat in it is challenging. 
After several experiments and analyzing the process IDs, the process whose dump eventually ended with chat messages was identified. We concluded that WhatsApp's process with the highest Process ID (PID) is the one that always contains the chat messages. The Algorithm is shown in Algorithm 1. The complexity of Algorithm 1 is of the order of O(n) where n is the number of processes running in the system. We identify the Process ID (PID) of that process to capture memory dump using the ProcDump tool and are able to retrieve all the user's chat messages. 

\begin{algorithm}
	\caption{Extraction of WhatsApp PID with the highest process ID} 
        \begin{algorithmic}
        \State $ProcessList=$List of WhatsApp process IDs
        \State $pid=$PID of WhatsApp Process
        \State $ProcessID=$PID of WhatsApp Process with highest process ID
        \State $ProcessList=[\ ]$
        \State $ProcessID=0$
		\For {each $process$ in $TaskManager$}
            \If {$process.processName='WhatsApp.exe'$}
                \State $ProcessList.push(process)$
                \If{$ProcessID<pid$}
                \State $ProcessID=pid$
            \EndIf
            \EndIf
        \EndFor
				\State \Return $ProcessID$
	\end{algorithmic} 
\end{algorithm}

\subsection{Extracting UNICODE \& ASCII Strings: Using Strings Tool}We used Microsoft's Sysinternals tool Strings v2.54\cite{strings} developed by Mark Russinovich to extract UNICODE and ASCII strings from captured memory dumps and saved them as a text file.

\subsection{Chat Messages Retrieval: Using Python Script} After extracting the textual file from the captured memory dump, we retrieved the chat messages involving only the sensitive keywords. We redirected the sentences that involved organizational sensitive keywords (Military data in this use case) for further sensitivity detection based on context to the Sensitive Data Detection Model (SDDM) as shown in Figure 1. The algorithmic implementation for retrieval of sensitive chat messages is described in Algorithm 2. The complexity of Algorithm 2 is of the order of O(s*l*m) where s= number of sensitive words, l = no. of lines in the complete text, and m=no.of extracted messages. 
\begin{algorithm}
	\caption{Retrieval of sensitive messages } 
	\begin{algorithmic}
	    \State $\bold{Output}:$AllMessage storing the sensitive messages  
	    \State $\bold{Variables}:te.txt=$stores all messages`
	    \State $pid=$PID of WhatsApp Process
	    \State $s\_word=$List of all Sensitive Words
	    \State $CompleteText=tostring($list of all extracted messages from te.txt$)$
	    \State $AllMessage=[\ ]$
		\For {each $sensitiveword$ in $s\_word$}
		    \State $data=CompleteText$
		    \While {$len(data)>=len(sensitive word)$}
			    \State $i=$index of the first occurrence of sensitive word
                \If {$i=-1$}
                    \State break
                \EndIf
                \State $j=$(find first occurrence of \textbackslash n index in backward)+1
                \State $k=$(find first occurrence of \textbackslash n index in forward)-1
                \State $chat=$string from index j to index k
                \State $flag=0$
                \While {message extracted till now}
                    \If{chat match with a message more then 90\%}
                        \State $flag=1$
                    \EndIf
                \EndWhile
                \If{$flag=0$}
                    \State $AllMessage.append(chat)$
                \EndIf
                \State $data=data-chat$
            \EndWhile
		\EndFor
	\end{algorithmic} 
\end{algorithm}
   
\subsection{Context-based Sensitive Data Detection Model}

\par Our proposed method uses context-based feature extraction for efficient data detection. The user’s private data and the sensitive military data used in this proposed architecture use case are publicly unavailable. We have used a semi-supervised pre-trained module called BERT to identify sensitive data by generating vectors of data on contextual bases rather than keywords. 
\section{ Experiments, Results and Analysis }
We designed our experimental setup to test the proposed model. We deployed the proposed architecture on Windows 10 user machine and installed the WhatsApp desktop version. The aim of the experiment is to access the proposed architecture when the malicious insider communicates military sensitive data over Whatsapp. In the following subsections, we describe the experiments conducted to demonstrate the deployment of our proposed model for detecting insider data leakage.   
\subsection{Assumptions}
While conducting experiments, we assume that the user’s machine is not compromised, root access is not available to the user, the process memory dump of the IM application installed on the user’s computer is genuine and not affected by any malicious activity and the analysis doesn’t affect the system working. It is assumed that the insider uses a device where the solution is installed.

\subsection{Tools Utilised}
The following tools are utilized while designing the proposed privacy-aware memory forensics model for insider threat detection. 
\begin{itemize}
    \item Memory Dumping Tool: Microsoft ProcDump v10.11
    \item Strings extraction from Memory Image: Microsoft Strings V2.54
    \item Extraction of chat messages from Strings: Python Script
    \item Implementation of Model: Python Script 
\end{itemize}
\subsection{Implementation and Testing} 
We implemented our privacy-aware memory forensics architecture on a user machine, with the specifications given in Table II, to detect sensitive data leaks over WhatsApp Desktop app installed on a Windows Machine.

 Many experiments are conducted on this model to assess the performance by supplying the testing data set.
 The method proceeded in the following sequence:
\begin{itemize}
    \item Model identified the running WhatsApp process ID from system information. 
    \item Memory dump of that particular process captured by using PrcoDump command-line utility.
    \item Converted the memory dump file into readable form by retrieving the \textit{UNICODE}, \textit{ASCII} strings using the Strings tool and saved as a text file.
    \item Chat messages containing sensitive military keywords are retrieved along with the user’s mobile number from the text file of the memory dump.
    \item Identified the sensitivity of these retrieved chat messages using SDDM. 
    \item Alert Message is initiated consisting of a leaked sensitive message and user mobile number to the security center if any data breach is identified.  
\end{itemize}

\begin{table}[htbp]
\caption{Specification of User Machine}
\begin{center}
     %\caption{Specification of User Machine}
    \begin{tabular}{|c|c|c|}
       \hline
       OS  & Windows 10 \\
       \hline
       Processor  &  Intel(R) Core(TM) i5-10500 CPU @ 3.10GHz   3.10 GHz   \\
       \hline
       RAM &  32GB(31.8 GB usable)    \\
       \hline 
       System Type & 64-bit Operating System, x64-based processor \\
       \hline 
    \end{tabular}
    \vspace{2mm}

%\end{tabular}
%\label{tab1}
\end{center}
\end{table}

\begin{table*}[hbt!]
\caption{Comparison of proposed architecture with previous work }\label{tab1}
\begin{center}
\resizebox{\textwidth}{!}{\begin{tabular}{l c c c c c}
\hline

\textbf{Author} &  \textbf{Live Memory } & \textbf{Privacy}  & \textbf{Lightweight} & \textbf{Context} & \textbf{Implemented}\\
 
\hline
Srinivasan \cite{policy} & ${\times}$ &$\times$ & $\times$ & NA & $\times$ \\
\hline
Waleed Halboob et al.\cite{quarter}   & ${\times}$ & ${\times}$ & ${\times}$ & NA & ${\times}$ \\

\hline
M Burmester et al \cite{choose}  & ${\times}$ & ${\times}$ & ${\times}$ & ${\times}$ & \checkmark \\

\hline
A Nieto, R Rios \cite{Psurvey} & ${\times}$ & ${\times}$ & ${\times}$ & NA & ${\times}$ \\

\hline
Ali Dehghantanha et al\cite{privacy}    &${\times}$ & ${\times}$ & ${\times}$ & NA & ${\times}$ \\

\hline
Padmavathi Iyer et al\cite{email}  & \checkmark & ${\times}$ &${\times}$ &${\times}$ & \checkmark \\
 
\hline
Abdullah Kazim et al\cite{chat}   & \checkmark & ${\times}$ & ${\times}$ & ${\times}$ & \checkmark \\
 
\hline
Al-Saleh et al \cite{network}  & \checkmark & ${\times}$ & ${\times}$ & NA & \checkmark \\

\hline
Gaurav et al \cite{doh} & \checkmark & ${\times}$ & ${\times}$ & ${\times}$ & \checkmark \\
\hline
Proposed work & \checkmark & \checkmark & \checkmark & \checkmark & \checkmark \\
\hline
\end{tabular}}

\vspace{2mm}
{Note: \checkmark - Yes, ${\times}$ - No, NA - not addressed  }
\end{center}
\end{table*}

\subsection{Results and Analysis}
\par An offline test on the BERT model indicates that the model performed the task with 95\% accuracy, and the corresponding confusion matrix is shown in Figure 2. We have used \textit{Recevier Operating Characteristic (ROC) }for evaluating the performance of our model.
In our case,
\begin{center}
True Positive Rate (TPR) $TPR = 138/(138+12)= 0.92$ and \end{center}
\begin{center}
True Negative Rate (TNR) $TNR = 148/(148+2)= 0.98$ \end{center}
Thus the accuracy of our model is 95\%
\par We tested the trained model by supplying some sensitive and normal messages as input, and the model identified the sensitive messages based on the context and generated the corresponding output values; during the experiments, it was observed that the model provided a sensitivity score above 0.5 for sensitive messages and less than 0.5 for user personal messages.
\begin{figure}
\centering
    \includegraphics[width=.30\textwidth]{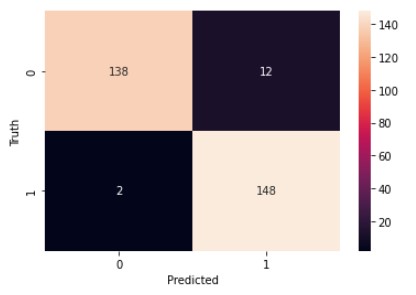}
    \caption{Performance of BERT model on test data (1: Sensitive Data; 0: Normal Data)}
    \label{fig:5}  
\end{figure}

\par The proposed architecture extracted the chat messages containing sensitive data from the memory dump and identified all the sensitive messages exchanged through WhatsApp desktop application based on the context. It can be clearly seen in Figure 3.

\begin{figure*}
    
    \includegraphics[width=.95\textwidth]{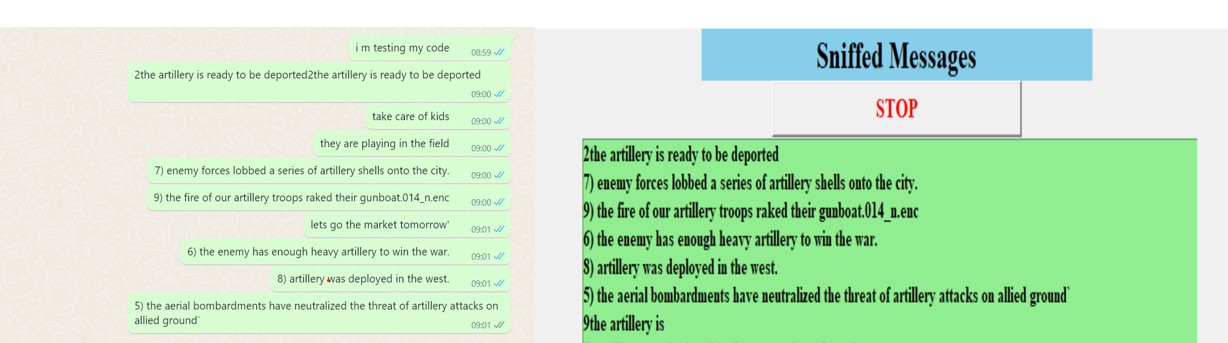}
    \caption{Captured Sensitive Messages}
    \label{fig:}
    
\end{figure*}
The private messages exchanged are not detected, maintaining the user's privacy with memory forensics. This is the scenario where organizations provide IT infrastructure to employees like desktops and laptops in security-sensitive environments. Our tool will run in the background over a host machine, capturing context-sensitive data.

During experiments, we observed that deleted messages from the hard disk are also recovered from the RAM. The proposed model is lightweight and thus significantly reduces the computational power. It captures the process memory instead of the entire RAM. Hence, the data that needs to be analyzed is in Kilo Bytes instead of some Giga Bytes. The processing time is also reduced accordingly. Our experimental results demonstrate the design feasibility of detecting malicious insider threats by using privacy-aware memory forensics techniques. 
\par The performance of the proposed privacy-aware memory forensics architecture while addressing the research gaps is compared with previous research work and tabulated in Table III. It is evident from the table that \begin{itemize}
\item only 50\% of the schemes work on live memory forensics. The significance of using live memory is that the data is available in raw form i,e it is not encrypted. Thus it gives a lot of scope to analyze the data. 
\item None of the schemes implements user privacy preservation during memory forensics. All the schemes capture the entire RAM where the user's private data is also available. Using Memory Forensics on the entire RAM breaches users' privacy.
\item None of the schemes works on per process capturing of RAM. Working with the entire captured RAM is computationally expensive compared to per-process captured RAM. 
\item Most of the models perform keyword-based searches instead of context-based ones. Keyword-based searches are recommended for identifying the required data but since no context-based searches are done, it leads to the detection of user's private messages as well. Thus, context-based search is required so that the user's privacy is maintained.
\end{itemize} 
Our proposed model only targets specific applications and applies privacy preservation on live memory forensics using the context-based approach, thus addressing all the identified gaps.

\section{Conclusions, Limitations and Future Scope}
We demonstrated optimal data leakage detection application of privacy-aware memory forensics in the military use case. All research gaps identified earlier are addressed by our model. Our model inspects process-based live memory instead of the entire RAM reducing the computations and hence is lightweight. The proposed method works for the new versions of operating systems because of the utilization of Microsoft utilities for acquiring memory (ProcDump) and converting dump files (Strings tool) in contrast to the third-party open-source tools which may not function properly. User privacy is ensured as only the sensitive data is determined from the live memory dump as our model mounts a context-based search for sensitive words which ensures the detection of only sensitive data. The privacy preservation feature of the proposed novel method escalates the memory forensics capabilities in the field of information security. Our model launched on the user machine, functions all actions automatically and initiates an alert message with necessary data to the security center.  
\par Since our model is designed for the military use case, one limitation of our scheme was the limited data set for training the model.
\par Interesting future work in this direction is to implement privacy preservation at the kernel API level so that any memory acquisition tool cannot access the user’s private data.

\end{document}